\def\ptitle{\tiny Energy bounds for a class of singular potentials and some related series}
\font\tr=cmr12                          
\font\bf=cmbx12                         
\font\it=cmti12                         
\font\trbig=cmbx12 scaled 1500          
\font\tiny=cmr10                        
\output={\shipout\vbox{\makeheadline
                                      \ifnum\the\pageno>1 {\hrule}  \fi 
                                      {\pagebody}   
                                      \makefootline}
                   \advancepageno}

\headline{\noindent {\ifnum\the\pageno>1 
                                   {\tiny \ptitle\hfil
page~\the\pageno}\fi}}
\footline{}

\tr 
\def\ni{\noindent}             

\def\hi#1#2{$#1$\kern -2pt-#2} 
\def\hy#1#2{#1-\kern -2pt$#2$} 

\def\ppl#1{{\noindent\leftskip 9 cm #1\vskip 0 pt}} 

\def\dbox#1{\hbox{\vrule 
\vbox{\hrule \vskip #1\hbox{\hskip #1\vbox{\hsize=#1}\hskip #1}\vskip #1 
\hrule}\vrule}} 
\def\qed{\hfill \dbox{0.05true in}} 
\baselineskip 15 true pt  
\parskip=0pt plus 5pt 
\parindent 0.25in
\hsize 6.0 true in 
\hoffset 0.25 true in 
\emergencystretch=0.6 in                 
\vfuzz 0.4 in                            
\hfuzz  0.4 in                           
\vglue 0.1true in
\mathsurround=2pt                        
\topskip=24pt                            
\newcount\zz  \zz=0  
\newcount\q   
\newcount\qq    \qq=0  

\def\pref#1#2#3#4#5{\frenchspacing \global \advance \q by 1     
    \edef#1{\the\q}{\ifnum \zz=1{\item{$^{\the\q}$}{#2}{\bf #3},{ #4.}{~#5}\medskip} \fi}}

\def\bref #1#2#3#4#5{\frenchspacing \global \advance \q by 1     
    \edef#1{\the\q}
    {\ifnum \zz=1 { %
       \item{$^{\the\q}$} 
       {#2}, {\it #3} {(#4).}{~#5}\medskip} \fi}}

\def\gref #1#2{\frenchspacing \global \advance \q by 1  
    \edef#1{\the\q}
    {\ifnum \zz=1 { %
       \item{$^{\the\q}$} 
       {#2.}\medskip} \fi}}

 \def\sref #1{[#1]}

\def\references#1{\zz=#1
   \parskip=2pt plus 1pt   
   {\ifnum \zz=1 {\noindent \bf References \medskip} \fi} \q=\qq
\pref{\harr}{E. M. Harrell, Ann. Phys. }{105}{ 379-406 (1977)}{Sec. 3, proposition 3.2}
\pref{\an}
{V. C. Aguilera-Navarro and R. Guardiola, J. Math. Phys. }{32}{2135-2141 (1991)}{ Sec. 2, Eq. 7.}
\pref{\eb}
{E. S. Est\'evez-Bret\'on and G. A. Est\'evez-Bret\'on, J. Math. Phys. }{34}{437-440 (1993)}{Sec. 2, Eq. 17}
\pref{\zn}
{M. Znojil, J. Math. Phys. }{34} {4914 (1993)}{}    
\bref{\grr}
{George E. Andrews, Richard Askey and Ranjan Roy }{Special Functions}{Cambridge (1999)}{p. 66}
\pref{\hala}{R. Hall, N. Saad and A. von Keviczky, J. Math. Phys. }{39}{6345-6351 (1998)}{}
\pref{\halb}{R. Hall and N. Saad, J. Phys. A: Math. Gen. }{33}{569-578 (2000)}{}
\pref{\halc}{R. Hall and N. Saad, J. Phys. A: Math. Gen. }{33}{5531-5537 (2000)}{Sec. 2, Eq.(2.9)}
\pref{\hald}{R. Hall, N. Saad and A. von Keviczky, J. Phys. A: Math. Gen. }{34}{1169-1179 (2001)}{}
\pref{\hale}{R. Hall, N. Saad and A. von Keviczky, J. Math. Phys. }{43}{94-112 (2002)}{}
\bref{\luke}
{Yudell L. Luke} {The special Functions and their Approximations, Vol. I, }{Academic press (1969)}{p. 26}
\pref{\bcf}
{B. L. Burrows, M. Cohen and Tova Feldmann, J. Phys. A: Math. Gen. }{20}{ 889-897 (1987)}{}
\bref{\luks}
{Yudell L. Luke} {The special Functions and their Approximations, Vol. I, }{Academic press (1969)}{Ch. III, Sec 3.3}

\pref{\muod}{Omar Mustafa and Maen Odeh, J. Phys. B: At. Mol. Opt. Phys. }{32} {3055-3063 (1999)  }{}
\pref{\muod}{Omar Mustafa and Maen Odeh, J. Phys. A: Math. Gen. }{33} {5207-5217 (2000)}{}
\gref{\jski}{J. Skibi\'nski, e-print quant-ph/0007059}
\pref{\zno}
{M. Znojil, Physics Letters A }{164}{ 138 (1992)}{}
\pref{\znoj}
{M. Znojil, J. Math. Phys. }{30}{ 23-27 (1989)}{}
\pref{\nag}
{N. Nag and R. Roychoudhury, Czechoslovak J. Phys. }{46}{ 343-351 (1996)}{}
\pref{\roy}
{M. Znojil and R. Roychoudhury, Czechoslovak J. Phys. }{48}{ 1-8 (1998)}{}
\pref{\anlk}
{V. C. Aguilera-Navarro and E. Ley Koo, Int. J. Theor. Phys. }{36}{157-166 (1997)}{}
\pref{\pease}
{P. Chang and Chen-Shiung Hsue, Phys. Rev. A }{49}{4448-4456 (1994)}{}
\pref{\acn}
{V. C. Aguilera-Navarro,A. L. Coelho and N. Uttah, Phys. Rev. A }{49}{1477-1479 (1994)}{}
\pref{\halh}
{R. Hall and N. Saad, Canad. J. Phys. }{73}{493-496 (1995)}{}
\bref{\messiah}
{Albert Messiah} {Quantum Mechanics, }{Dover (1999), Vol. II}{Chapter XVI, p. 685}
\bref{\land}
{L. D. Landau and E. M. Lifshitz} {Quantum Mechanics: Non-relativistic Theory}{Pergamon Press(1977)}{Chapter VI, especially Eq.(38.6), Eq.(38.10), and problem (2)}
\pref{\half}
{R. L. Hall, N. Saad and A. von Keviczky, J. Phys. A: Math. Gen. } {34}{11287 (2001)}{}
\pref{\halg}
{R. L. Hall and N. Saad, J. Phys. A: Math. Gen. } {33} {5531-3337 (2002)}{}
\bref{\luk}
{Yudell L. Luke} {The special Functions and their Approximations, Vol. I }{Academic press (1969)}{p. 110, formula (36)}

\pref{\oktay}{Oktay Sinano\v glu, Phys. Rev. }{112}{491-492 and 493-499(1961)}{Especially, footnote 6 in the second article.}
\bref{\mf}
{P.M. Morse and H. Feshbach} {Methods of Theoretical Physics}{McGraw-Hill Book Comapny (1953)}{pp. 1120}
\pref{\kill}{J. Killingbeck, Rep. Prog. Phys. }{40}{963-1031 (1977)}{Sec. 3.5, especially the discussion below Eq. 3.34}
\pref{\per}
{L\" owdin Per-Olov, J. Math. Phys. }{3}{969-982 (1962)}{p. 979}

\pref{\znojil}
{M. Znojil, F. Gemperle and Omar Mustafa, J. Phys. A: Math. Gen. }{35}{5781-5793 (2002)}{}

}

\references{0}    
\ppl{CUQM-95}\ppl{math-ph/0211058} 
\ppl{November 2002}\medskip 
\vskip 0.8 true in
\centerline{\bf\trbig Energy bounds for a class of singular potentials }
\centerline{\bf\trbig and some related series}
\medskip
\vskip 0.25 true in
\centerline{Nasser Saad$^\dagger$, Richard L. Hall$^\ddagger$,  and Attila B. von Keviczky$^\ddagger$}
\bigskip
{\leftskip=0pt plus 1fil
\rightskip=0pt plus 1fil\parfillskip=0pt
\obeylines
$^\dagger$Department of Mathematics and Computer Science,
University of Prince Edward Island, 
550 University Avenue, Charlottetown, 
PEI, Canada C1A 4P3.\par}

\medskip
{\leftskip=0pt plus 1fil
\rightskip=0pt plus 1fil\parfillskip=0pt
\obeylines
$^\ddagger$Department of Mathematics and Statistics, Concordia University,
1455 de Maisonneuve Boulevard West, Montr\'eal, 
Qu\'ebec, Canada H3G 1M8.\par}

\vskip 1 true in
\centerline{\bf Abstract}\bigskip
\noindent Perturbation expansions up to third order for the generalized spiked harmonic oscillator Hamiltonians  $H=-{d^2\over dx^2}+ x^2+{A\over x^2}+{\lambda\over x^\alpha},\ (A\geq 0,2\gamma > \alpha, \gamma=1+{1\over 2}\sqrt{1+4A}),$ and small values of the coupling $\lambda>0$, are developed. Upper and lower bounds for the eigenvalues are computed by means of the procedure of Burrows {\it et al} [J. Phys. A: Math. Gen. {\bf 20}, 889-897 (1987)] for assessing the accuracy of a truncated perturbation expansion.  Closed-form sums for some related perturbation double infinite series then immediately follow as a result of this investigation.
\bigskip
\vfil\eject
\ni{\bf 1. Introduction}
\medskip
It is well known that, although many perturbation expansions diverge, they may actually be asymptotic expansions whose first few terms can yield good approximations.
 The family of spiked harmonic oscillator Hamiltonians 
$$
H=H_0+\lambda V=-{d^2\over dx^2}+x^2+{\lambda\over x^\alpha}\quad (0\leq x<\infty)\eqno(1.1)
$$
affords interesting examples of this phenomenon. Harrell $\sref{\harr}$ have shown that the  familiar Rayleigh-Schr\"odinger perturbation series diverge according as $n \geq {1\over \alpha-2},$ where $n$ is the order of the Rayleigh-Schr\"odinger term. For example, the first-order perturbation correction diverges for $\alpha\geq 3$, while the second-order correction term diverges if $\alpha\geq {5\over 2},$ and so on. In a sequel of articles, Aguilera-Navarro et al $\sref{\an}$, Est\'vez-Bret\'on et al $\sref{\eb}$, and Znojil $\sref{\zn}$ have shown for the case of $\alpha<5/2$, the so called `non-singular' case, that the perturbation series of the ground-state energy up to the second-order corrections is given by 
$$
E(\lambda,\alpha)=3+{\Gamma({3-\alpha\over 2})\over \Gamma({3\over 2})}\lambda-{\Gamma^2({3-\alpha\over 2})\over \Gamma^2({3\over 2})}\sum\limits_{i=1}^\infty {({\alpha\over 2})_i^2\over 4~i~({3\over 2})_i~i!}\lambda^2+\dots. \quad\quad\hbox{ for }\alpha<5/2.\eqno(1.2) 
$$ 
Based on resummation techniques, an analysis of Aguilera-Navarro {\it et al} $\sref{\an}$  showed that
$$\eqalign{
\sum\limits_{i=1}^\infty {({\alpha\over 2})_i^2\over 4~i~({3\over 2})_i~i!}&=
\sum\limits_{i=1}^\infty {({\alpha\over 2})_i^2\over 4~(i+1)~({3\over 2})_i~i!}
+\sum\limits_{i=1}^\infty {({\alpha\over 2})_i^2\over 4~i~(i+1)~({3\over 2})_i~i!}
\cr
&={1\over 8({\alpha\over 2}-1)^2}\bigg[{}_2F_1({\alpha\over 2}-1,{\alpha\over 2}-1; {1\over 2};1)-1-2({\alpha\over 2}-1)^2\bigg]
+\sum\limits_{i=1}^\infty {({\alpha\over 2})_i^2\over 4~i~(i+1)~({3\over 2})_i~i!}},\eqno(1.3)
$$
where ${}_2F_1(a,b;c;z)$ is the known Gauss hypergeometric function $\sref{\grr}$ with circle of convergence $|z|=1$. For the limiting case $\alpha\rightarrow 2$, the first term on the right-hand side of (1.3) was shown by Est\'vez-Bret\'on {\it et al} $\sref{\eb}$ using l'H\^opital's rule to be
$$
\lim_{\alpha\rightarrow 2} {1\over 8({\alpha\over 2}-1)^2}\bigg[{}_2F_1({\alpha\over 2}-1,{\alpha\over 2}-1; {1\over 2};1)-1-2({\alpha\over 2}-1)^2\bigg]= {\pi^2\over 16}-{1\over 4}.\eqno(1.4)
$$
Znojil, soon afterwards $\sref{\zn}$, showed elegantly that (1.4) follows immediately by manipulating the Maclaurin expansion of the gamma function. Recently, Hall and Saad $\sref{\hala-\hale}$ investigated a larger class so called generalized spiked harmonic oscillator Hamiltonians
$$
H=H_0+\lambda V=-{d^2\over dx^2}+x^2+{A\over x^2}+{\lambda\over x^\alpha}\quad (A\geq 0).\eqno(1.5)
$$
The Gol'dman and Krivchenkov Hamiltonian $H_0=-{d^2\over dx^2}+x^2+{A\over x^2},$   which admits the exact solutions
$$\psi_n(x)=(-1)^n\sqrt{{2(\gamma)_n}\over n!\Gamma(\gamma)}x^{\gamma-{1\over 2}}e^{-{1\over 2}x^2}{}_1F_1(-n,\gamma,x^2),\eqno(1.6)
$$
with exact eigenenergies
$$E_n=4n+2\gamma,\quad n=0,1,2,\dots,\quad \gamma=1+{1\over 2}\sqrt{1+4A},\eqno(1.7)$$
is regarded as the unperturbed part, and the operator $V(x) =x^{-\alpha}$ as the perturbed part. They obtained $\sref{\halc}$ the energy expansion up to the second-order as 
$$
E(\lambda,\alpha)
= 2\gamma +{\Gamma(\gamma-{\alpha\over 2})\over \Gamma(\gamma)}\lambda-\lambda^2{\alpha^2\over 16\gamma}~{\Gamma^2(\gamma-{\alpha\over 2})\over \Gamma^2(\gamma)}{}_4F_3(1,1,1+{\alpha\over 2},1+{\alpha\over 2};2,2,\gamma+1;1)+\dots\eqno(1.8)
$$ valid for $\alpha<\gamma+1$, where $\gamma=1+{1\over 2}\sqrt{1+4A}.$
A closed form sum for the infinite series in (1.2) appears as especial case. In particular, for $\gamma=3/2$ or $A=0$, Eq.(1.8), for $\alpha<{5\over 2}$, reduces to
$$
E(\lambda,\alpha)
= 3 +{2\over \sqrt{\pi}}\Gamma({3-\alpha\over 2})\lambda-\lambda^2{\alpha^2\over 48}~{\Gamma^2({3-\alpha\over 2})\over \Gamma^2(\gamma)}{}_4F_3(1,1,1+{\alpha\over 2},1+{\alpha\over 2};2,2,{5\over 2};1)+\dots
$$ 
and closed-form sums of the infinite series in (1.2) follow immediately. Furthermore, for $\alpha =2,$ since 
$${}_4F_3(1,1,2,2;2,2,\gamma+1;1) = {}_2F_1(1,1;\gamma+1;1)={\Gamma(\gamma+1)\Gamma(\gamma-1)\over \Gamma(\gamma)\Gamma(\gamma)}={\gamma\over (\gamma-1)},$$
by means of Chu-Vandermonde theorem $\sref{\grr}$
$${}_2F_1(a,b;c;1)={\Gamma(c)\Gamma(c-a-b)\over \Gamma(c-a)\Gamma(c-b)},\quad \hbox{ for } c-a-b>0,\eqno(1.9)$$
the perturbation expansion (1.8) takes the very simple form
$$
E(\lambda,\alpha=2)=
2\gamma +{\lambda\over (\gamma-1)}-{\lambda^2\over 4(\gamma-1)^3}+\dots\eqno(1.10)
$$ 
This is obtained, as expected, by means of Taylor's expansion of the exact energy $2+\sqrt{1+4(A+\lambda)}$ about $\lambda=0$. In order to understand the result (1.4), however, we should note first
$$\eqalign{
\sum\limits_{i=1}^\infty {({\alpha\over 2})_i^2\over 4~(i+1)~(\gamma)_i~i!}&
={(\gamma-1)\over 4({\alpha\over 2}-1)^2}\bigg[{}_2F_1({\alpha\over 2}-1,{\alpha\over 2}-1;\gamma-1;1)-1-{({\alpha\over 2}-1)^2\over (\gamma-1)}\bigg]
\cr
&=
{(\gamma-1)\over 4({\alpha\over 2}-1)^2}
\bigg[{\Gamma(\gamma-1)\Gamma(\gamma-\alpha+1)\over \Gamma(\gamma-{\alpha\over 2})\Gamma(\gamma-{\alpha\over 2})}-1-{({\alpha\over 2}-1)^2\over (\gamma-1)}\bigg]\cr
&={(\gamma-1)\over 4({\alpha\over 2}-1)^2}
\bigg[{\Gamma(\gamma-1)\Gamma(\gamma-\alpha+1)\over \Gamma(\gamma-{\alpha\over 2})\Gamma(\gamma-{\alpha\over 2})}-1\bigg]-{1\over 4}
}
$$
where we have used (1.9). Now since
$$\lim_{\alpha\rightarrow 2}{1\over ({\alpha\over 2}-1)^2}
\bigg[{\Gamma(\gamma-1)\Gamma(\gamma-\alpha+1)\over \Gamma(\gamma-{\alpha\over 2})\Gamma(\gamma-{\alpha\over 2})}-1\bigg]=\psi^{(1)}(\gamma-1)
$$
we have
$$
\sum\limits_{i=1}^\infty {({1})_i^2\over (i+1)~(\gamma)_i~i!}={1\over 2\gamma} {}_3F_2(1,2,2;3,\gamma+1;1)=(\gamma-1)\psi^{(1)}(\gamma-1)-1\quad\hbox{ for } \gamma>1,\eqno(1.11)$$
where $\psi^{(1)}(z)$ is the first-derivative of the digamma function (or logarithmic derivative of the gamma function $\sref{\luke}$) . Further, since $\psi^{(1)}({1\over 2})={\pi^2\over 2}$, the result of (1.4) follows immediately by replacing $\gamma$ with $3/2$ in (1.11). 

The interesting feature of the expression (1.8) is that, it can be applied to the ground-state eigenenergy at the bottom of each angular-momentum subspace labelled by $l=0,1,2,\dots$ in \hi{N}{dimensions}: we just need to replace $A$ with $A\rightarrow A+(l+{1\over 2}(N-1))(l+{1\over 2}(N-3))$. Furthermore, as we shall prove in the next section, for $\alpha=4$ and $\gamma>3$ (or $A>3.75$), the perturbation expansion (1.8) takes the very simple form 
$$
E(\lambda,\alpha=4)= 2\gamma +{\lambda\over (\gamma-1)(\gamma-2)}-\lambda^2{4\gamma^2-15\gamma+13\over 4(\gamma-1)^3(\gamma-2)^3(\gamma-3)}+\dots,\eqno(1.12)
$$ 
where $\psi$ is the digamma function. For $\alpha=6$ and $\gamma>5$ (or $A>15.75$),  (1.8) becomes
$$\eqalign{
E(\lambda,\alpha=6)&=2\gamma+{\lambda\over (\gamma-1)(\gamma-2)(\gamma-3)}-
{\Gamma^2(\gamma-3)\over 8\Gamma^2(\gamma)}\bigg({( \gamma-2 )(\gamma-1 )\over (\gamma-5)(\gamma-4)}\cr
& +{2( \gamma-1 )\over ( \gamma-4)} + {40 - 57\gamma + 8\gamma^2 - \gamma^3\over ( \gamma-3 )( \gamma-2 )( \gamma-1 )}\bigg)\lambda^2+\dots.}\eqno(1.13)
$$ 
In Sec. 2 we shall extended these perturbation expansions to third-order corrections. In Sec. 3, we shall discuss upper and lower bounds for the eigenvalues by means of the procedure of Burrows {\it et al} $\sref{\bcf}$ for assessing the accuracy of a truncated perturbation expansion. These bounds will shed some light on the question regarding the acceleration of the variational method. Our conclusions and some remarks concerning the sums of some double infinite series will be given in Sec. 4. 

The functions  ${}_1F_{1}$ and ${}_4F_3$, mentioned above, are special cases of the generalized hypergeometric function ${\sref{\luks}}$ 
$$
{}_pF_{q}(\alpha_1,\alpha_2,\dots,\alpha_p;\beta_1,\beta_2,\dots,\beta_q;z)=\sum\limits_{k=0}^\infty 
{\prod\limits_{i=1}^p(\alpha_i)_k\over  \prod\limits_{j=1}^q(\beta_j)_k}{z^k\over k!},\eqno(1.14)
$$ 
where $p$ and $q$ are non-negative integers, and none of the $\beta_j,$ ($j=1,2,\dots,q$) is equal to zero or to a negative integer. If the series does not terminate (that is to say, none of the $\alpha_i$, $i=1,2,\dots,p$, is a negative integer), then the series, in the case $p=q+1$, converges or diverges accordingly as $|z|<1$ or $|z|>1$. For $z=1$, the series is convergent  provided
$
{\sum\limits_{j=1}^q \beta_j-\sum\limits_{i=1}^p \alpha_i}>0.
$ 
Here $(a)_n$, the shifted factorial (or {\it Pochhammer symbol}), is defined by
$$(a)_0=1,\quad (a)_n=a(a+1)(a+2)\dots (a+n-1)={\Gamma(a+n)\over \Gamma(a)},\quad n=1,2,\dots.\eqno(1.15)$$

\medskip
\ni{\bf 2. Third-order perturbation expansions}
\medskip
In this section we will expand the perturbation expansions (1.8) to the third-order correction. Although, we will concentrate on the cases of $\alpha=4$ and $\alpha=6,$ since they are the most relevant in the literature$\sref{\muod-\halh}$, for other values of $\alpha$ the procedure is similar. In order to lay the foundation of the perturbation expansion (1.8), we first review the Rayleigh-Schr\"odinger perturbation theory for a non-degenerate case $\sref{\messiah}$. The fundamental problem in perturbation theory is the solution of the Schr\"odinger equation $H\phi= E(\lambda)\phi$ when $H= H_0+\lambda V$.  The basic assumption is that $\phi$ and $E(\lambda)$ may be expanded in power series in the perturbation parameter $\lambda$:
$$\phi=\psi_0+\sum\limits_{i=1}^\infty \lambda^i\phi_i,\quad\quad E(\lambda)=E_0+\sum\limits_{i=1}^\infty \lambda^i \epsilon_i.\eqno(2.1)$$
Here $\psi_0$ is a solution to the unperturbed problem $H_0\psi_0=E_0\psi_0$. We also choose the normalization $(\psi_0,\phi)=1,$ which implies that the higher-order corrections $\phi_1,\phi_2\dots$ are orthogonal to $\psi_0$. Perturbation theory tells us in this case that 
$$\epsilon_1=(\psi_0,V\psi_0),\quad \epsilon_2=(\psi_0,V\phi_1),\quad \epsilon_3=(\phi_1,V\phi_1)-\epsilon_1(\phi_1,\phi_1),\quad\dots,\eqno(2.2)$$
or, equivalently$\sref{\land},$
$$\epsilon_1=(\psi_0,V\psi_0),~ \epsilon_2=\sum\limits_{i=1}^\infty {|V_{0i}|^2\over E_i-E_0},~ \epsilon_3=\sum\limits_{s=1}^\infty\sum\limits_{k=1}^\infty {V_{0s}V_{sk}V_{k0}\over (E_s-E_0)(E_k-E_0)}-\epsilon_1\sum\limits_{i=1}^\infty {|V_{0i}|^2\over (E_i-E_0)^2},~\dots\eqno(2.3)$$
From (2.2) it is clear that the first-order wave function $\phi_1$ determines the energy to the third-order.  The matrix elements $V_{ij} = (\psi_i,V\psi_j)$ in (2.3) are computed by means of the basis solution $\{\psi_n\}$ of the unperturbed Hamiltonian $H_0$. For the generalized spiked harmonic oscillator Hamiltonian
(1.5), the expectation values of the operator $V(x) = x^{-\alpha}$ with respect to the Gol'dman and Krivchenkov basis (1.6) are given explicitly by
$$
V_{ij}=(-1)^{i+j} 
{({\alpha\over 2})_i \Gamma(\gamma-{\alpha\over 2})\over (\gamma)_i\Gamma(\gamma)}
\sqrt{{(\gamma)_i(\gamma)_j\over i!j!}}
{}_3F_{2}(-j,\gamma-{\alpha\over 2},
1-{\alpha\over 2};\gamma,1-i-{\alpha\over 2};1).\eqno(2.4)
$$
Of particular interest is
$$
V_{i0}=V_{0i}= (-1)^{i} 
{({\alpha\over 2})_i \over (\gamma)_i}
\sqrt{{(\gamma)_i\over i!}}{\Gamma(\gamma-{\alpha\over 2})\over \Gamma(\gamma)}.
\eqno(2.5)
$$
Recently, Hall {\it et al} $\sref{\half-\halg}$ have shown that the first-order correction of the wavefunction, in the case of $\alpha=2$, is given by
$$
\phi_1(x)=
{1\over \sqrt{2}}
{x^{\gamma-{1\over 2}}e^{-{x^2\over 2}}\over (\gamma-1)\sqrt{\Gamma(\gamma)}}
\bigg[\log(x)-{1\over 2}\psi(\gamma)\bigg],\quad \hbox{for } \gamma>1.\eqno(2.6)
$$
Therefore from (2.2) and (2.3), by using (2.5), we have
$$-{\Gamma^2(\gamma-1)\over \Gamma^2(\gamma)}\sum\limits_{i=1}^\infty {(1)_i^2\over 4~i~({\gamma})_i~i!}=\int\limits_0^\infty x^{-2}\psi_0(x)\phi_1(x)dx=-{1\over 4(\gamma-1)^3},$$
as shown previously using summation technique. This idea can be used to obtain a simple form by expressing ${}_4F_3$ in (1.8) in terms of elementary functions.  These indeed are facilitated by the closed expression of the first-order correction of the wave functions developed earlier $\sref{\half-\halg}$. In the case $\alpha=4$, the first-order correction of the wave function reads 
$$
\phi_1(x)=
{1\over 2\sqrt{2}}
{x^{\gamma-{1\over 2}}e^{-{x^2\over 2}}\over (\gamma-2)(\gamma-1)\sqrt{\Gamma(\gamma)}}
\bigg[\log(x^2)-\psi(\gamma)-{\gamma-1\over x^2}+1\bigg],\quad \hbox{for } \gamma>2.\eqno(2.7)
$$
where $\psi$ is the digamma function $\sref{\luke}$. Using (2.2) and (2.3), we have 
$${}_4F_3(1,1,3,3;2,2,\gamma+1;1)={1\over 4}{\gamma(4\gamma^2-15\gamma+13)\over (\gamma-1)(\gamma-2)(\gamma-3)},\quad \gamma >3$$
and therefore the perturbation expansion (1.12) follows immediately. These particular values of ${}_4F_3(1,1,3,3;2,2,\gamma+1;1)$ can be verified by means of the following lemma that extends the earlier identity 
$${}_3F_2(a,b,c+1;d,c;z)={}_2F_1(a,b;d;z)+{ab\over cd}~z~{}_2F_1(a+1,b+1;d+1;z)$$
given by Luke $\sref{\luk}$. The proof follows immediately by use of the series representation for the hypergeometric functions ${}_3F_2$ and ${}_2F_1,$ as given by (1.14).
\medskip
\noindent {\bf Lemma 1:} {\it For $|z|<1$,
$$\eqalign{
{}_4F_3(a,b,c+1,d+1;e,c,d;z)&={}_2F_1(a,b;e;z)+{ab\over e c}(1+{c+1\over d})z~{}_2F_1(a+1,b+1;e+1;z)\cr
&+{(a)_2(b)_2\over dc (e)_2}z^2{}_2F_1(a+2,b+2;e+2;z)}\eqno(2.8)$$
Further, in the case of $|z|=1$ and $e-a-b>2$,
$$\eqalign{
{}_4F_3(a,b,c+1,d+1;e,c,d;1)&={\Gamma(e)\over \Gamma(e-a)\Gamma(e-b)}\bigg[\Gamma(e-a-b)+{ab\over c}(1+{c+1\over d})\Gamma(e-a-b-1)\cr
&+{(a)_2(b)_2\over dc }\Gamma(e-a-b-2)\bigg]}\eqno(2.9)$$
}

\ni In the case of $\alpha = 6$, the first-order correction of the wave function reads $\sref{\half-\halg}$
$$
\phi_1(x)={1\over 2\sqrt{2}}{\Gamma(\gamma-3)\over \Gamma(\gamma)\sqrt{\Gamma(\gamma)}}x^{\gamma-1/2}e^{-x^2/2}\bigg[\log(x^2)-\psi(\gamma)+ {3\over 2}-{\gamma-1\over x^2}-{(\gamma-1)(\gamma-2)\over 2x^4}\bigg]\eqno(2.10)
$$
consequently, from $\epsilon_2=(\psi_0,x^{-6}\phi_1)$, we have for $\gamma>5$
$$
{}_4F_3(1,1,4,4;2,2,\gamma+1;1)={\gamma\over 18}\bigg({( \gamma-2 )(\gamma-1 )\over (\gamma-5)(\gamma-4)}+{2( \gamma-1 )\over ( \gamma-4)} + {(40 - 57\gamma +24\gamma^2 -3 \gamma^3)\over ( \gamma-3 )( \gamma-2 )( \gamma-1 )}\bigg).\eqno(2.11)
$$
Therfore Eq.(1.8) takes the simpler form (1.13), as the result of (2.11). 
In order to extend (1.12) and (1.13) to the third-order perturbation correction, we need only use the expression $\epsilon_3=(\phi_1,V\phi_1)-\epsilon_1(\phi_1,\phi_1),$ as mentioned in (2.2). Before we proceed with our calculations we shall first prove the following general result concerning the first-order correction of the wave function.
\medskip 
\noindent{\bf Lemma 2:} {\it The first-order perturbation correction $\phi_1(x)$ of the exact solution of Hamiltonian (1.5), with arbitrary $\alpha$, satisfies the following normalization condition 
$$(\phi_1,\phi_1)= {\alpha^2\over 64\gamma}{\Gamma^2(\gamma-{\alpha\over 2})\over  \Gamma^2(\gamma)}~{}_5F_4(1,1,1,{\alpha\over 2}+1,{\alpha\over 2}+1;2,2,2,\gamma+1;1)$$
as long as $\alpha<\gamma+2$.
}

\noindent{PROOF:} We note that, by comparing the expression for $\epsilon_3$ in (2.2) and (2.3), we find
$$(\phi_1,\phi_1)=\sum\limits_{i=1}^\infty {|V_{0i}|^2\over (E_i-E_0)^2}.$$
For the Hamiltonian (1.5), $V_{0i}$ is given by (2.5) and $E_i$ is given by (1.7); therefore we have
$$\eqalign{(\phi_1,\phi_1)&={1\over 16} {\Gamma^2(\gamma-{\alpha\over 2})\over  \Gamma^2(\gamma)} \sum\limits_{i=1}^\infty{({\alpha\over 2})_i^2\over i^2~i!~(\gamma)_i}\cr
&={1\over 16} {\Gamma^2(\gamma-{\alpha\over 2})\over  \Gamma^2(\gamma)} \sum\limits_{i=0}^\infty{({\alpha\over 2})_{i+1}^2\over (i+1)^2~(i+1)!~(\gamma)_{i+1}}\cr
&={\alpha^2\over 64\gamma} {\Gamma^2(\gamma-{\alpha\over 2})\over  \Gamma^2(\gamma)} \sum\limits_{i=0}^\infty{(1)_i~(1)_i~ (1)_i~({\alpha\over 2}+1)_{i}^2\over (2)_i~(2)_i~(2)_i~(\gamma+1)_{i}~}{1\over i!}\cr
&={\alpha^2\over 64\gamma}{\Gamma^2(\gamma-{\alpha\over 2})\over  \Gamma^2(\gamma)}~{}_5F_4(1,1,1,{\alpha\over 2}+1,{\alpha\over 2}+1;2,2,2,\gamma+1;1)
}
$$
where we have used the Pochhammer identities $(a)_{n+1}=a(a+1)_n$, $(1)_n=n!$ and $(2)_n=(n+1)!$  (see (1.15)), and the series representation for the hypergeometric function ${}_5F_4,$ as given by (1.14).\qed 
 
Direct computations, using $\epsilon_3=(\phi_1,x^{-4}\phi_1)-\epsilon_1(\phi_1,\phi_1)$ where $\phi_1$ is given by (2.7) and $\epsilon_1={\Gamma(\gamma-2)\over \Gamma(\gamma)}$ leads, for $\alpha=4$ and $\gamma>4$, to
$$\eqalign{
E(\lambda,\alpha=4)=& 2\gamma +{\lambda\over (\gamma-1)(\gamma-2)}-\lambda^2{4\gamma^2-15\gamma+13\over 4(\gamma-1)^3(\gamma-2)^3(\gamma-3)}\cr
&+\bigg\{{16\gamma^5-175\gamma^4+742\gamma^3-1525\gamma^2+1520\gamma-590\over 8(\gamma-4)(\gamma-3)^2(\gamma-2)^5(\gamma-1)^5}\bigg\}\lambda^3+\dots
}\eqno(2.12)
$$  
For the case of $\alpha=6$, the first-order correction of the wavefunction is given by (2.10). After
some straightforward algebraic calculations, the ground-state perturbation expansion, up to the third-order of $\lambda$ and valid for $\gamma>7,$  now reads
$$
E(\lambda,\alpha=6)=2\gamma+\epsilon_1\lambda+\epsilon_2\lambda^2+\epsilon_3\lambda^3+\dots,\eqno(2.13)
$$
where
$$ \epsilon_1={\lambda\over (\gamma-1)(\gamma-2)(\gamma-3)},
$$
$$
\epsilon_2=-{\Gamma^2(-3 + \gamma)\over 8\Gamma^2(\gamma)}\bigg({( \gamma-2 )(\gamma-1 )\over (\gamma-5)(\gamma-4)} +{2( \gamma-1 )\over ( \gamma-4)} + {(40 - 3\gamma(19 + (-8 + \gamma) \gamma))\over ( \gamma-3 )( \gamma-2 )( \gamma-1 )}\bigg),
$$
and
$\epsilon_3={I_1\over I_2}$, 
for
$$\eqalign{
&
I_1=
192088 - 655905\gamma + 945811\gamma^2 - 751923\gamma^3 + 360811\gamma^4 - 107151\gamma^5 + 19257\gamma^6 - 1917\gamma^7 + 81\gamma^8,\cr
&I_2= 8(\gamma-7)(\gamma-5)^2(\gamma-4)(\gamma-3)^5(\gamma-2)^5(\gamma-1)^5.
}
$$

A first reading of the articles by Sinano\v glu $\sref{\oktay}$ (the main results of which are not affected by his false claim), or even the work of Morse and Feshbach $\sref{\mf}$ on perturbation theory, one understands that the expressions (2.12) and (2.13) are upper bounds to the exact energy since all the odd-order energies would form upper bounds to the exact energy. This is not in fact true because $\epsilon_2$ in the general perturbation expansion (2.1) will always have a negative sign, thus not guaraneeing the upper bounds $\sref{\kill-\per}$.  However, it {\it is} possible to obtain a definite upper bound to the exact eigenvalue by means of the perturbation expansion. Thus
$$
E(\lambda,\alpha)=E_0+\epsilon_1\lambda+{\epsilon_2\lambda^2+\epsilon_3\lambda^3\over 1+\lambda^2(\phi_1,\phi_1)},\eqno(2.14)
$$
where $(\phi_1,\phi_1)$ is given by Lemma 2. The upper bound (2.14) can easily be demonstrated by applying the variational principle to the approximate wave function $\phi=\psi_0+\lambda\phi_1,$ where $\psi_0$ and $\phi_1$ satisfies the zero- and first-order perturbation equations
$$H_0\psi_0=E_0\psi_0,\quad (H_0-E_0)\phi_1=(E_1-V)\psi_0.\eqno(2.15)$$

In Table (1), we compare the upper bounds obtained by means of (2.14) in the case of $\alpha =4$ and those of Aguilera-Navarro and Koo obtained by variational analysis using appropriate trial functions. In this next section, we shall obtain the symmetric lower and upper bound by means of the method of Burrows {\it et al} $\sref{\bcf}$.
\medskip
\ni{\bf 3. Lower and upper bounds}
\medskip
It is natural to ask: how small $\lambda$ should be for the perturbation expansions (2.12) and (2.13) to be valid?.  The question can be answered by studying upper and lower bounds to the eigenvalues. Based on the difference between the bounds we can infer a definite indication of the accuracy of truncated Rayleigh-Schr\"odinger perturbation series, such as (2.12) and (2.13). Wide bounds show that the truncated Rayleigh-Schr\"odinger perturbation series is suspect, while tight bounds demonstrate the high accuracy of the truncated expansion. For our purposes, the most suitable procedure developed for assessing the accuracy of a truncated perturbation expansion is due to Burrows {\it et al}$\sref{\bcf}$. A brief review of the method is presented here: for further details the reader is referred to the original article. Most derivations of bounds for eigenvalues of self-adjoint operators start from a consideration of positive definite function
$$
(\mu(\phi,\epsilon), \mu(\phi,\epsilon))=([H-\epsilon]\phi,[H-\epsilon]\phi)=(H\phi,H\phi)-(\phi,H\phi)^2+(\epsilon-(\phi,H\phi))^2\geq 0,\eqno(3.1)
$$
where $H$ is the operator in question, $\epsilon$ is a positive parameter, and $\phi$ is a suitably chosen (normalized) function. If we expand the normalized function $\phi$ in terms of the complete set of eigenfunctions $\{\phi_n\}$ of $H$ with eigenvalues $E_n(\lambda)$, $\phi=\sum_{n}a_n\phi_n$, $a_n=(\phi,\phi_n)$,  $(\phi,\phi)=1=\sum_{n}a_n^2$, we can express the positive definite function in (3.1) as
$$
(\mu(\phi,\epsilon), \mu(\phi,\epsilon))=\sum_{n}a_n^2(E_n(\lambda)-\epsilon)^2\geq 0
$$
Let us assume that we have picked the value of $\epsilon$ to lie closest to the value of the $i$th eigenvalue $E_i$, i.e.
$$
(\mu(\phi,\epsilon), \mu(\phi,\epsilon))=\sum_{n}a_n^2(E_n(\lambda)-\epsilon)^2\geq (E_i(\lambda)-\epsilon)^2 \geq 0\eqno(3.2)
$$
Combining (2.6) and (2.7), it can easily be seen that
$$
f_{-}(\epsilon)\leq E_i(\lambda)\leq f_{+}(\epsilon),\eqno(3.3a)
$$
where
$$f_{\pm}(\epsilon) = \epsilon\pm \sqrt{\parallel H\phi\parallel^2 -(\phi,H\phi)^2+(\epsilon-(\phi,H\phi))^2}.\eqno(3.3b)
$$
It is not hard to show that $f_{\pm}(\epsilon)$ is indeed a monotonic increasing function of $\epsilon.$ This result will turn out to be useful in the following discussion.
The bounds of Burrows {\it et al} follow $\sref{\bcf}$ by setting 
$$
\mu(\phi,E_p(\lambda))=[H_0+\lambda V-E_p(\lambda)]\phi\eqno(3.4)
$$
where 
$$
\phi=N_1(\psi_0+\lambda\phi_1),\quad E_p(\lambda)=E_{0}+\sum_{i=1}^p\lambda^i \epsilon_{i}
\eqno(3.5)
$$
and, for all $p\leq 3$, $\psi_0$ and $\phi_1$ satisfy the zero- and first-order equations of the Rayleigh-Sch\"odinger perturbation theory (2.15). Further, the $\epsilon_{i},i=1,2,3$ are given by means of (2.2). Here, 
$N_1$ in Eq.(3.5) is a normalization constant for the truncated first-order expansion of the exact wavefunction:
$$N_1=(1+\lambda^2(\phi_1,\phi_1))^{-1/2}.$$
 If $\phi$ and $E_p(\lambda)$ were exact, $\mu=0.$ Thus we expect $\mu$ to be small if $\phi$ and $E_p(\lambda)$ are good approximations to the exact solutions.  Consequently, a good test of the approximations (3.3a-b) may be made by examining the value of the norm $\parallel \mu \parallel=\sqrt{\mu^2}$.
Simple calculations, using (3.5) and (2.2), now give
$$
\parallel \mu(\phi,E_1(\lambda)) \parallel=N_1\lambda^2 (\phi_1, (V-\epsilon_{1})^2\phi_1)^{1/2},\eqno(3.6)
$$
$$
\parallel \mu(\phi,E_2(\lambda)) \parallel=\{\parallel \eta(\phi, E_1(\lambda)) \parallel^2+\lambda^4 \epsilon_2\{\epsilon_2-2N_1^2(\epsilon_{2}+\lambda \epsilon_3)\}\}^{1/2},\eqno(3.7)
$$
and
$$
\parallel \mu(\phi,E_3(\lambda)) \parallel=\{\parallel \eta(\phi,E_2(\lambda)) \parallel^2+\lambda^4\{2\lambda^3 \epsilon_2\epsilon_3(1-N_1^2)-\lambda^2(1+\lambda^2)N_1^2\epsilon_3^2+\lambda^4\epsilon_3^2\} \}^{1/2}\eqno(3.8)
$$
where we have re-produced the  formulas of Burrows {\it et al} $\sref{\bcf}$ for computational convenience. In this case, (3.3a) implies
$$
f_{-}(E_p(\lambda)) \leq E(\lambda)\leq  f_+(E_p(\lambda))\eqno(3.9a)
$$
where 
$$f_{\pm}(E_p(\lambda))=E_p(\lambda)~\pm \parallel \mu(\phi,E_p(\lambda)) \parallel\quad\hbox{ for } p=1,2,3.\eqno(3.9b)$$
The only new integral (beyond the usual integrals of Rayleigh-Schr\"odinger perturbation series) is seen to be $(\phi_1|(V-\epsilon_1)^2\phi_1)$ which restricts the value of $\gamma$, for example in case of $\alpha=4$, to be greater than $4$ even if we have used the first-order approximation $\epsilon_1$ (for which $\gamma>2$ is sufficient).  This is, of course, due to the bound's dependence on $\epsilon_3$ which required $\gamma>4.$ The result in this case, however, is very useful $\sref{\acn,\znojil}$ when the radial Schr\"odinger equation is characterized  by {\it large} angular momenta $l$. For $\gamma= 4.5~ (i.e.~ A=12$ or $l=3$ for $A=l(l+1)$) and $\lambda = 0.001$, the first-order perturbation correction yields $9.000~114~285$ with an error bound of $\pm 4.8346\times 10^{-8}$. The second-order perturbation corrections yields $9.000~114~279$ with error bounds of $\pm 4.7879\times 10^{-8}$; while $\epsilon_3$ yields $9.000~114~279$ with an upper bound of $9.000~114~327$ and a lower bound of $9.000~114~231$. Now, for any fixed $\phi$, the bounding functions $f_{\pm}(E_p(\lambda))$ are easily shown to be monotonic increasing functions of $E_p(\lambda)$, $p=1,2,3$, as we indicated above.  Consequently the optimal bound for the set $\{E_1(\lambda)=E_0+\lambda\epsilon_1,E_2(\lambda)=E_0+\lambda\epsilon_1+\lambda^2\epsilon_2, E_3(\lambda)=E_0+\lambda\epsilon_1+\lambda^2\epsilon_2+\lambda^3\epsilon_3\}$ is indeed given, for $\lambda< {|\epsilon_2|\over \epsilon_3}$, by
$$
f_{-}(E_1(\lambda))\leq E(\lambda)\leq f_{+}(E_2(\lambda)).\eqno(3.10)
$$
The inequality $\lambda< {|\epsilon_2|\over \epsilon_3}$ allows us to order the approximated eigenvalues as $E_1(\lambda)> E_3(\lambda)> E_2(\lambda),$ for the sign of $\epsilon_2$ is always negative and the sign of $\epsilon_3$ is positive for moderate values of $\lambda$. In Table II we have verified these results by obtaining upper and lower bounds for the eigenvalues by means of (3.6-8); underlined values are the optimal bounds. Similar bounds can be obtained for the case of $\alpha = 6$ by using (2.13). Although, the upper bounds obtained by this method are less accurate than the upper bounds obtained by means of (2.14), the advantage of this method is the symmetric lower and upper bounds avaliable through (3.9).
\medskip
\ni{\bf 4. Conclusions and some remarks}
\medskip
The main results of the present article are concrete upper- and lower-bound formulas (2.14),  (3.9), and (3.10). There are many variational methods avaliable to solve the eigenvalue problem for the Hamiltonian (1.5), however they provide only upper bounds and usually no information is avaliable concerning the accuracy of the method other than comparison with numerical solutions of the Schr\"odinger equation in question. Furthermore, for very small values of the parameter $\lambda,$ variational methods are usually slow and a large number of the matrix elements are needed to obtain sufficient accuracy. We have presented upper and lower bounds for such situations which, as table (I) and (II) indicate, provide excellent results for very small values of $\lambda$.  Although, the techniques used to produce the present results are standard, the ability of these techniques to generate explicit bounds is a consequence of our pervious achievements, yielding concrete forms for the first-order perturbation corrections of the wave functions.    

Aside from the upper and lower bounds obtained, there are also some interesting results concerning a closed-form sums for double infinite series that follow directly from the present work. It is clear from (2.2) and (2.3) that
$$ \sum\limits_{m=1}^\infty\sum\limits_{n=1}^\infty {V_{0n}V_{nm}V_{m0}\over (E_0-E_n)(E_0-E_m)}=(\phi_1,V\phi_1)\eqno(4.1)$$
where $V_{nm}, n=1,2,\dots,m=1,2,\dots$ are given by (2.4). We will now look at the cases $\alpha=2,4,6,\dots.$ Similar results can be obtained for the cases of $\alpha =1,3,5,\dots$ by means of the first-order corrections for the wave functions given previously $\sref{\half-\halg};$ however, the calculations will be more involved for such cases. For $\alpha=2$, we know that the matrix elements (i.e from (2.4)) read
$$
V_{nm}=\cases{(-1)^{n+m}
{\Gamma(\gamma-1)\over \Gamma(\gamma)}
\sqrt{m!(\gamma)_n\over n!(\gamma)_m}& if $m\geq n$,\cr
\ \cr
(-1)^{n+m}
{\Gamma(\gamma-1)\over {\Gamma(\gamma)}}
\sqrt{n!(\gamma)_m\over m!(\gamma)_n}& if $n\geq m$.\cr}
\eqno(4.2)
$$
On other hand, the first-order correction of the wave function in this case reads $\sref{\half-\halg}$ 
$$
\phi_1(x)=
{1\over \sqrt{2}}
{x^{\gamma-{1\over 2}}e^{-{x^2\over 2}}\over (\gamma-1)\sqrt{\Gamma(\gamma)}}
\bigg[\log(x)-{1\over 2}\psi(\gamma)\bigg],\quad \hbox{for } \gamma>1.\eqno(4.3)
$$
Consequently, the following results follow immediately,
\medskip 
\noindent{\bf Lemma 3.} {\it For $\gamma>1$ and $V_{nm}$ as given by (4.2), we have
$$\sum\limits_{m=1}^\infty\sum\limits_{n=1}^\infty {V_{0n}V_{nm}V_{m0}\over 16nm}=
{1\over 8(\gamma-1)^5}+{\psi^{(1)}(\gamma)\over 16(\gamma-1)^3},\eqno(4.4)
$$
where
$\psi^{(1)}(\gamma)$ is the first derivative of the digamma functions.}

The proof of this Lemma is obtained by calculating the inner product of the right-hand side of (4.1) by means of (4.3) for $0\leq x<\infty$, where $V(x)=x^{-2}$ and $E_n=4n+2\gamma$ ($n=0,1,2,\dots$). For the case $\alpha = 4$ and $\gamma>2$, the matrix elements (2.4) read
$$
V_{nm}=\cases{
(-1)^{n+m}
{\Gamma(\gamma-2)\over \Gamma(\gamma+1)}
\sqrt{m!(\gamma)_n\over n!(\gamma)_m}[\gamma(m-n+1)+2n]& if $m\geq n$,\cr
\ \cr
(-1)^{n+m}
{\Gamma(\gamma-2)\over \Gamma(\gamma+1)}
\sqrt{n!(\gamma)_m\over m!(\gamma)_n}[\gamma(n-m+1)+2m]& if $n\geq m$.\cr}
\eqno(4.5)
$$
On the other hand, the first-order corrections of the wave function for this case are given by (2.7). Therefore (4.1) leads to the following results
\medskip 
\noindent{\bf Lemma 4.} {\it For $\gamma>4$ and $V_{nm}$ as given by (4.5), we have
$$\sum\limits_{m=1}^\infty\sum\limits_{n=1}^\infty {V_{0n}V_{nm}V_{m0}\over 16nm}=
{-820+1954\gamma-1753\gamma^2+694\gamma^3-90\gamma^4-12\gamma^5+3\gamma^6\over 16(\gamma-4)(\gamma-3)^2(\gamma-2)^5(\gamma-1)^5}+{\psi^{(1)}(\gamma)\over 16(\gamma-2)^3(\gamma-1)^3},\eqno(4.6)
$$
where
$\psi^{(1)}(\gamma)$ is the first derivative of the digamma functions.}
\medskip
As final case that we illustrate, namely $\alpha=6$ and $\gamma>3$, we point to the fact that Eq.(2.4) lets us deduce   
$$
V_{nm}=\cases{
(-1)^{n+m}
{\Gamma(\gamma-3)\over 2\Gamma(\gamma+2)}
\sqrt{{m! (\gamma)_n}\over n!(\gamma)_m}\times\cr[(2+m)(1+m)\gamma(\gamma+1)-2n(1+m)
(\gamma-3)(\gamma+1)-n(1-n)(\gamma-2)(\gamma-3)]& if $m\geq n$,\cr
\ \cr
(-1)^{n+m}
{\Gamma(\gamma-3)\over 2\Gamma(\gamma+2)}
\sqrt{n!(\gamma)_m\over m!(\gamma)_n}\times\cr[(2+n)(1+n)\gamma(\gamma+1)-2m(1+n)(\gamma-3)(\gamma+1)-m(1-m)(\gamma-2)(\gamma-3)]& if $n\geq m$.\cr}
\eqno(4.7)
$$
where the first order correction for the wave function is now given by (2.10). Therefore, by means of (4.1), we conclude
\medskip 
\noindent{\bf Lemma 5.} {\it For $\gamma>7$ and $V_{nm}$ as given by (4.7), we have
$$\sum\limits_{m=1}^\infty\sum\limits_{n=1}^\infty {V_{0n}V_{nm}V_{m0}\over 16nm}={I_1\over I_2}+{\psi^{(1)}\over 16(\gamma-3)^3(\gamma-2)^3(\gamma-1)^3},\eqno(4.8)$$
where
$$I_1=522652-1717440\gamma+2371931\gamma^2-1785046\gamma^3+792061\gamma^4-206964\gamma^5+28725\gamma^6-1158\gamma^7-169\gamma^8+16\gamma^9,
$$
and
$$I_2=
32(\gamma-7)(\gamma-5)^2(\gamma-4)(\gamma-3)^5(\gamma-2)^5(\gamma-1)^5,
$$
where
$\psi^{(1)}(\gamma)$ is the first derivative of the digamma function.}

\bigskip
\noindent {\bf Acknowledgments}
\medskip Partial financial support of this work under Grants GP3438 and GP249507
from the Natural Sciences and Engineering Research Council of Canada is gratefully 
acknowledged by two of us, respectively [RLH] and [NS].

\vfil\eject
\noindent {\bf Table (I)}~~~A comparison between the upper bounds for the Hamiltonian (1.5), for a wide range of values of $A=l(l+1)$ and $\lambda$, by formula (2.14) and the bounds $E_a^U$ obtained by Aguilera-Navaro {\it et al} $\sref{\anlk}$. Exact results $E$ found by direct numerical solution of Sch\"odinger's equation are also presented.

\bigskip
\vbox{\tabskip=0pt\offinterlineskip
\def\tablerule{\noalign{\hrule}}
\def\vr{\vrule height 12pt}
\halign to420pt{\strut#\vr&#
\tabskip=1em plus2em
&\hfil#\hfil
&\vrule#
&\hfil#\hfil
&\vrule#
&\hfil#\hfil
&\vrule#
&\hfil#\hfil
&\vrule#
&\hfil#\hfil
&\vr#\tabskip=0pt\cr
\tablerule&&$\lambda$&&$l$&&$E_a^U$&&$E^{U}$&&$E$&\cr\tablerule
&&$0.001$&&$3$&&$~9.000~114~279~82$&&$9.000~114~279~12$&&$9.000~114~279~12$&\cr
&&~&&$4$&&$11.000~063~490~8$&&$11.000~063~490~7$&&$11.000~063~490~74$&\cr
&&~&&$5$&&$13.000~040~403~7$&&$13.000~040~403~6$&&$13.000~040~403~64$&\cr\tablerule
&&$0.01$&&$3$&&$~9.001~142~268~25$&&$9.001~142~199~48$&&$9.001~142~199~40$&\cr
&&~&&$4$&&$11.000~634~795~5$&&$11.000~634~788~8$&&$11.000~634~788~89$&\cr
&&~&&$5$&&$13.000~404~001~8$&&$13.000~404~000~6$&&$13.000~404~000~60$&\cr\tablerule
&&$0.1$&&$3$&&$~9.011~370~328~09$&&$9.011~364~261~69$&&$9.011~364~026~18$&\cr
&&~&&$4$&&$11.006~336~739~4$&&$11.006~336~100~1$&&$11.006~336~099~23$&\cr
&&~&&$5$&&$13.004~036~546~4$&&$13.004~036~432~5$&&$13.004~036~432~52$&\cr\tablerule
&&$1$&&$3$&&$~9.109~013~250~38$&&$9.109~311~262~10$&&$9.108~658~607~52$&\cr
&&~&&$4$&&$11.062~293~143~4$&&$11.062~249~282~0$&&$11.062~241~719~38$&\cr
&&~&&$5$&&$13.040~025~483~8$&&$13.040~015~551~5$&&$13.040~015~183~06$&\cr
&&~&&$50$&&$103.000~400~037$&&$103.000~400~036$&&$103.000~400~036~76$&\cr
\tablerule
}}\bigskip\medskip

\noindent {\bf  Table (II)}~~~Eigenvalue Bounds for different values of $\lambda$ for the Hamiltonian $H=-{{d^2}\over {dx^2}}+x^2+{12\over x^{2}}+{\lambda\over x^{4}}$. The underlined values are the optimal bounds according to inequality (3.10).
\bigskip
\noindent\hfil\vbox{%
\offinterlineskip
\tabskip=0pt
\halign{\tabskip=3pt
\vrule#\strut&#\strut\hfil&\vrule#\strut&\hfil#\strut\hfil&\vrule#\strut&\hfil#\strut\hfil&\vrule#\strut&\hfil#\strut\hfil&\vrule#\strut&\hfil#\strut\hfil&\vrule#\strut&\hfil#\strut\hfil&\vrule#\strut&\hfil#\strut\hfil&\vrule#\strut\tabskip=0pt\cr
\multispan2&\multispan4{\hrulefill}&\multispan4{\hrulefill}&\multispan4\hrulefill\cr
\multispan2&\multispan4\vrule\hfil$\epsilon_1$\hfil&&\multispan3\hfil$\epsilon_2$\hfil&&\multispan3\hfil $\epsilon_3$\hfil&\cr
\multispan5&\multispan5&\omit&\omit\vrule\cr\noalign{\hrule}
\multispan2\vrule~$\lambda$&&$E_L$&&
		$E^U$&&$E_L$&&$E^U$&&$E_L$&&$E^U$&\cr
\noalign{\hrule}
&$0.001$&&\underbar{$9.000114234$}&&$9.000114334$&&$9.000114231$&&\underbar{$9.000114327$}&&$9.000114231$&&$9.000114327$ &\cr
\noalign{\hrule}
&$0.01$&&\underbar{$9.001138022$}&&$9.001147691$&&$9.001137408$&&\underbar{$9.001146987$}&&$9.001137409$&&$9.001146989$ &\cr
\noalign{\hrule}
&$0.1$&&\underbar{$9.010945111$}&&$9.011912031$&&$9.010883476$&&\underbar{$9.011841809$}&&$9.010885097$&&$9.011843425$ &\cr
\noalign{\hrule}
&$1$&&\underbar{$9.065963521$}&&$9.162607906$&&$9.059599522$&&\underbar{$9.155786288$}&&$9.061245282$&&$9.157377241$ &\cr
\noalign{\hrule}}
}
\vfil\eject
\references{1}

\end